# Classification of Boolean Functions where Affine Functions are Uniformly Distributed

Ranjeet Kumar Rout, Pabitra Pal Choudhury[1], Sudhakar Sahoo[2]

[1]Applied Statistics Unit, Indian Statistical Institute, Kolkata, India

[2]Institute of Mathematics & Applications, Bhubaneswar, India

Email: ranjeetkumarrout@gmail.com, pabitrapalchoudhury@gmail.com, sudhakar.sahoo@gmail.com

**Abstract:** Classification of Non-linear Boolean functions is a long-standing problem in the area of theoretical computer science. In this paper, effort has been made to achieve a systematic classification of all n-variable Boolean functions, where only one affine Boolean function belongs to each class. Two different methods are proposed to achieve this classification. The first method is a recursive procedure that uses the Cartesian product of sets starting from the set of 1-variable Boolean function and in the second method classification is achieved through a set of invariant bit positions with respect to an affine function belonging to that class. The invariant bit positions also provide information concerning the size and symmetry properties of the classes/sub-classes, such that the members of classes/sub-classes satisfy certain similar properties.

**Key words:** Affine Boolean Function, Truth Table, Classification, Carry Value Transformation.

## 1. Introduction

The classification of Boolean functions having a representative in each class is meaningful for the following two reasons: first, equivalent functions in each class possess similar properties and second, the number of representatives is much less than the number of Boolean functions. In the past, two Boolean functions of n-variables are considered equivalent and belongs to the same logic family (or equivalence classes), if they differ only by permutation or complementation of their variables. The formula for counting the number of such equivalence classes is given in [2]. Further, given the class decomposition, how to select a *representative assembly,* with one member from each family is explained in [1]. The linear group and the affine Boolean function group of transformations are defined and an algorithm is proposed for counting the number of classes under both groups [3]. The set of n-input functions is classified according to three criteria: the number of functions, the number of P classes and the number of NPN classes in [4]. Classification of the affine equivalence classes of cosets of the first order Reed-Muller code with respect to cryptographic properties such as correlation immunity, resiliency and propagation characteristics are discussed in [5]. Three Variable Boolean functions in the name of 3-neighbourhood Cellular Automata rules are classified on the basis of Hamming distance with respect to linear rules **[**7**]**. The characterization of 3-variable non-linear Boolean functions in the name of 1-D cellular automata rules is under taken in three different ways by Boolean derivatives [6], by deviant states [7] and by matrices [8].

In this paper we propose two methods for generating equivalence classes of Boolean functions. The first one is a recursive approach to classify n-variable Boolean functions starting from 1-variable to higher variables. The second method, on keeping some of the bit positions fixed with respect to an affine Boolean function and varying other bit position in it generates exactly the same classes of n-variable Boolean functions.

The arrangement of the paper is as follows: in section 2, the literature of Boolean functions of different variables relevant to our work is reviewed; in section 3, the method of recursive classification of n-variable Boolean function is introduced and we discuss the properties of these classes. Based on these properties another efficient method proposed for generating the same classes of n-variable Boolean functions; in section 4, we studied the behavior of those classes by using different binary operations such as Hamming Distance, XOR operation and Carry value transformation(CVT); section 5 is the conclusion.

## 2. Relevant Review

An n-variable Boolean function $f(x_1, x_2, x_3 \ldots x_n)$ is a mapping from the set of all possible n-bit strings denoted by $\{0,1\}^n$ into the alphabet $\{0,1\}$. For n- variables the number of Boolean function is $2^{2^n}$ and it is also represented as the output column of its truth table denoted by $f_R^n$, which is a binary string of length $2^n$. $f_R^n$ is also known as the function number $R$ in n –variable [9]. Where $R$ is the decimal equivalent of the binary sequence (starting from bottom to top) of the function in the truth table and obeys the Wolfram's naming convention [11]. The compliment of $f_R^n$ is denoted as $\overline{f_R^n}$.

A Boolean function of algebraic degree at most unity is called an affine Boolean function, the general form for n-variable affine function is

$$f_{affine}(x_1, x_2, \ldots, x_n) = k_n x_n \oplus k_{n-1} x_{n-1} \oplus \ldots \oplus k_2 x_2 \oplus k_1 x_1 \oplus k_0, k_i \in \{0,1\} \text{ and } \forall i \in \{0,1,2,3,\ldots,n\}$$

If the constant term $k_0$ of an affine function is zero then the function is called a *linear Boolean function*. Thus, *affine Boolean functions* are either linear Boolean functions or their compliments. For n binary variables, the total number of *affine Boolean function* is $2^{n+1}$ out of which $2^n$ are linear. The 16 *affine Boolean functions* for 3-variable are **0, 60, 90, 102, 150, 170, 204, 240, 15, 51, 85, 105, 153, 165, 195, 255** out of which first eight are linear Boolean functions and rest Boolean functions are their corresponding compliments [3].

The concatenation of the Boolean function $f$ with itself and the concatenation of $f$ with its complement $\bar{f}$ are denoted as $ff$ and $f\bar{f}$. For example, if $f = \begin{pmatrix} 0 \\ 0 \end{pmatrix}$, then $ff = \begin{pmatrix} 0 \\ 0 \\ 0 \\ 0 \end{pmatrix}$ and $f\bar{f} = \begin{pmatrix} 0 \\ 0 \\ 1 \\ 1 \end{pmatrix}$.

Note that if $f$ is a Boolean function in n-variable then $ff$ and $f\bar{f}$ are Boolean functions of n+1-variable.

**Theorem 1:** $f$ **is linear if and only if** $ff$ **and** $f\bar{f}$ **are linear.** [9]

That is Linear Boolean function concatenated with non-linear Boolean function is again Linear Boolean function provided the non–linear Boolean function is the complement with which it is concatenated and Linear Boolean function concatenated with itself is a linear Boolean function [9].Rest of all concatenations gives new non-linear Boolean functions.

**Corollary 1:** $f$ **is an affine Boolean function if and only if** $ff, f\bar{f}, \bar{f}f$ **and** $\bar{f}\bar{f}$ are affine Boolean functions.

**Proof:** It can be proved on using Theorem 1 and the definition of affine Boolean function.

## 3. Proposed Methods of Classification of Boolean functions

In this section two different methods are proposed to classify the set of all possible n-variable Boolean functions of same cardinality and each class containing a single affine function.

## 3.1 A Recursive procedure to classify the n-variable Boolean function

Let $S = \{\{00\}, \{10\}, \{11\}, \{01\}\}$ be a set of all 1-variable Boolean functions, where all the Boolean functions are *affine*. Let $S' = \{\{00\}, \{10\}\}$ be a set containing all linear Boolean functions of 1-variable and $S'' = \{\{11\}, \{01\}\}$ is a set that contains the complements of the set $S'$. The Cartesian product of the sets $S$ with $S'$ and $S''$ is defined as follows

$$S \times S' = \{\{\mathbf{0000}, 0010\}, \{1000, \mathbf{1010}\}, \{\mathbf{1100}, 1110\}, \{0100, \mathbf{0110}\}\} --- (1)$$

and

$$S \times S'' = \{\{\mathbf{0011}, 0001\}, \{1011, \mathbf{1001}\}, \{\mathbf{1111}, 1101\}, \{0111, \mathbf{0101}\}\} --- (2)$$

Note that S contains four classes each containing a 1-variable Boolean function where as the set $(S \times S') \cup (S \times S'')$ contains eight disjoint classes that exhaust all 2-variable Boolean functions and each class contains exactly one affine Boolean function which are highlighted.

This process is repeated, that is $(S \times S') \cup (S \times S'')$ now becomes $S$, $(S \times S')$ becomes $S'$ and $(S \times S'')$ becomes $S''$ for next higher variable. This recursive procedure, which repeated up to (n-1) times, classify the set off all n-variable Boolean functions into $2^{n+1}$ number of disjoint classes having a property that each class contains exactly one n-variable affine Boolean function along with some n-variable non linear Boolean functions.

It is because of the fact that $(S \times S') \cup (S \times S'') = S \times (S' \cup S'') = S \times S = set\ of\ all\ n$-variable Boolean functions and $(S \times S') \cap (S \times S'') = S \times (S' \cap S'') = S \times \phi = \phi$. And also each class contains exactly one n-variable affine Boolean function can be ascertained using **Corollary 1** of section 2.

**Illustration: (from 2-variable classes to 3-variable classes)**
From equation 1 and 2 and using the above procedure now the set

$$S = \begin{Bmatrix} \{\mathbf{0000}, 0010\}, \{1000, \mathbf{1010}\}, \{\mathbf{1100}, 1110\}, \{0100, \mathbf{0110}\} \\ \{\mathbf{0011}, 0001\}, \{1011, \mathbf{1001}\}, \{\mathbf{1111}, 1101\}, \{0111, \mathbf{0101}\} \end{Bmatrix}$$ which contains all 2-variable

Boolean functions of 2-variable and $S' = \{\{\mathbf{0000}, 0010\}, \{1000, \mathbf{1010}\}, \{\mathbf{1100}, 1110\}, \{0100, \mathbf{0110}\}\}$ is the first four classes of $S$ and $S'' = \{\{\mathbf{0011}, 0001\}, \{1011, \mathbf{1001}\}, \{\mathbf{1111}, 1101\}, \{0111, \mathbf{0101}\}\}$ is the set containing the rest classes of $S$ which is the complements of the set $S'$. Now the classes of 3-variable are calculated as follows:

$$S \times S' = \left\{ \begin{pmatrix} \mathbf{00000000}, \\ 00000010, \\ 00001000, \\ 00001010, \\ 00001100, \\ 00001110, \\ 00000100, \\ 00000110, \\ 00100000, \\ 00100010, \\ 00101000, \\ 00101010, \\ 00101100, \\ 00101110, \\ 00100100, \\ 00100110 \end{pmatrix}, \{CLASS\ 2\}, \dots, \{CLASS\ 8\} \right\}, \quad S \times S'' = \left\{ \begin{pmatrix} 00000011, \\ 00000001, \\ 00001011, \\ 00001001, \\ \mathbf{00001111}, \\ 00001101, \\ 00000111, \\ 00000101, \\ 00100011, \\ 00100001, \\ 00101011, \\ 00101001, \\ 00101111, \\ 00101101, \\ 00100111, \\ 00100101 \end{pmatrix}, \{CLASS\ 10\}, \dots, \{CLASS\ 16\} \right\}$$

The naming of the classes are given as Class 1, Class 2,…, and Class $2^{n+1}$ such that the compliment of Class $k$ is the Class $(2^n + k)$ where $k = 1, 2, 3, \dots, 2^{n+1}$. In the above figure only the members of CLASS 1 and CLASS 13 are shown and other classes of Boolean functions are shown in **Appendix-I**.

**Theorem 1:** The number of classes in the above classification is the number of Affine Boolean function in n-variables i.e. $2^{n+1}$.

**Proof:** As each class contains exactly one affine Boolean function, hence the number classes of n-variable are $2^{n+1}$.

**Theorem 2:** The classes are of equal size and the cardinality of each class equals to $2^{2^n-(n+1)}$.

**Proof:** The equal size of the classes easily follows from the cardinality of the two sets $(S \times S')$ and $(S \times S'')$. On using **Theorem 1** the cardinality of each class

$$=\frac{total\,number\,of\,n-variable\,Boolean\,functions}{total\,number\,of\,n-variable\,affine\,Boolean\,functions} = \frac{2^{2^n}}{2^{n+1}} = 2^{2^n-(n+1)}.$$

**Theorem 3:** All the Boolean functions of n-variable of Class k $for\ k = 1, 2, 3, \ldots, 2^n\ are\ even$ where as the Boolean functions in Class $(2^n + k)\ are\ odd.$

**Proof:** From the beginning when n=1 the set $S'$ contains all even Boolean functions and the set $S''$ contains all odd Boolean function and hence the recursive procedure using the Cartesian product also preserve the same properties.

Interestingly the relation defined in the recursive procedure is operating on the set of (n-1)-variable Boolean functions but the partitioned obtained in the set of n-variable Boolean functions. This motivates us to find out an equivalence relation on the set of n-variable Boolean functions which divides it into disjoint equivalence classes.

**Theorem 4:** For each class of n-variable the length of a Boolean function is $2^n$, out of which (n+1) bits are fixed and $2^n - (n+1)$ bits are changing with respect to an affine Boolean function of that class.

1. The bit positions of a Boolean function which are fixed in a class are calculated using the following formula:
   $2^0, 2^0 + 2^{n-1}, 2^0 + 2^{n-1} + 2^{n-2}, 2^0 + 2^{n-1} + 2^{n-2} + 2^{n-3}, \ldots \ldots, 2^0 + 2^{n-1} + \cdots + 2^{n-n}$.
2. The bit positions of a Boolean function which are changed in a class for n-variable are calculated using the formula is given below:
   $[1, 2, 3, 4, \ldots \ldots \ldots 2^{n-1} - 1] + [2^0], [1, 2, 3, 4, \ldots \ldots \ldots 2^{n-2} - 1] + [2^0 + 2^{n-1}], [1, 2, 3, 4, \ldots 2^{n-3} - 1] + [2^0 + 2^{n-1} + 2^{n-3}], \ldots \ldots \ldots [2^{n-n}] + [2^0 + 2^{n-1} + \cdots + 2^2]$.

**Proof:** It can be proved using Corollary 1 and the method of induction starting from the one-variable Boolean functions to higher-variable Boolean functions.

**Illustration:**

1. For 1-variable Boolean function all the bit positions are fixed and the bit positions are $2^0 = 1$ and $2^0 + 2^{1-1} = 2$. Similarly, for 2-variable Boolean function three bit positions are fixed and the bit positions are $2^0 = 1$, $2^0 + 2^{2-1} = 3$ and $2^0 + 2^{2-1} + 2^{2-2} = 4$. For 3-variable Boolean function four bit positions are fixed and the bit positions are $2^0 = 1$, $2^0 + 2^{3-1} = 5$, $2^0 + 2^{3-1} + 2^{3-2} = 7$ and $2^0 + 2^{3-1} + 2^{3-2} + 2^{3-3} = 8$ (ref. Appendix I).

2. For 1-varible Boolean function no bit positions are changing. But, for 2-varible Boolean function one bit position is changed and the bit position is $[1] + [2^0] = 2$. Similarly, for 3-varible Boolean function four bit positions are changed and the bit positions are $[1, 2, 3] + [2^0] = 2, 3, 4$ and $[1] + [2^0 + 2^{3-1}] = 6$ (ref. Appendix I).

**Corollary 2:** The bit positions which are fixed or changed are invariant for all classes with respect to the affine Boolean function of n-variable.

**Proof:** As the formula given in **Theorem 4** is used to calculate the bit positions which are fixed or changed are same for all classes so the result follows.

Using the result of **Theorem 4,** an equivalence relation is defined on the set of all possible n-variable Boolean functions which is the basis for obtaining the same classes without using recursion. This relation is defined as follows:

Let $f$ and $g$ be two n-variable Boolean functions and $R$ is a binary relation defined on the set of all possible n-variable Boolean functions as follows, $f\,R\,g$ iff *"there exist (n+1) bit positions calculated using **Theorem 4**, where the values in those positions are same for the functions f and g"* . Clearly,

1. $f R f \quad \forall f$ , So, $R$ is reflexive .
2. If $f R g$ then $g R f$, So, $R$ is symmetric .
3. If $f R g$ and $g R h$ then $f R h$, So, $R$ is transitive.

 *Hence, R is an equivalence relation.*

The next procedure uses the above equivalence relation and generates the same classes efficiently in comparison with the recursive procedure of **Section 3.1**.

### 3.2 Procedure to generate the same class without using recursion

Let '$f_k^n$' is an affine Boolean function of n-variable and $B$ is an array which is used to keep the bit positions those are fixed for a Boolean function with respect to $f_k^n$. The output array B can be calculated by invoking the following **function** in an **algorithm** whose worst case time complexity is $O(n)$.

*Fixed-bit-Positions* $(f_k^n)$

1. Initialize $X = 2^n$
2. $for\, i = 0\, to\, n$
3. {
4. $B[i] = X$
5. $X = B[i] - 2^i$
6. }
7. return B

To get other non- linear Boolean functions in a class, one has to put all possible binary sequences of length $2^n - (n + 1)$ , except these fixed position values of $f_k^n$.

*The inferences drawn from the above classification method are as follows:*

1. The method of keeping some of the bit positions fixed and varying other bit positions with respect to a Boolean function shall be a handle to find out equivalence classes of same cardinality.
2. Number of equivalence classes is equals to $2^k$ where $k$ be the number of fixed positions.
3. The number of members in a particular class is $2^l$ for $0 \leq l \leq 2^n - k$ is the number of changing positions.
4. By changing the fixed position we can change the members of the classes.
5. How to select the representative functions from the set of all possible Boolean functions that decompose into disjoint equivalence classes of equal cardinality? The generators are all possible $k$ bit sequences in the fixed positions and rest of the positions are either 0/1(Don't care) can be a generator for the class. The number of generators for the proposed classification is $2^k$.
6. Given the class decomposition, any Boolean function of our proposed class including the affine can be a representative of that family. But taking the affine functions as representatives is the guarantee of being a member in every class.

## 4. Different Operations in Classes

In this section, classes are divided into several sub-classes using the Hamming Distance (HD) between the Boolean functions and the affine function in that class. Also the classes are analysed on performing XOR and CVT operations among the functions of a class.

### 4.1 Sub-Classification

Hamming Distance between two Boolean functions is denoted as $HD(f,g) = k$, where $k = 0, 1, 2, \ldots, 2^n - (n+1)$ where $f$ is a Boolean function and $g$ is an affine Boolean function and both belongs to same class of n-variable. Further, Boolean functions in a class having $HD = k$ with respect to the corresponding affine Boolean function forms sub-classes whose cardinality are **Binomial Coefficients of the form $2^n - (n+1)_{C_k}$** where $k = 0, 1, 2, \ldots, 2^n - (n+1)$.

**Illustration:** Table 4.1 shows the Boolean functions belonging to Class 1 in 3-variables where $f_0^3 = 00000000$ is the affine Boolean function. There are five sub-classes having cardinality 1, 4, 6, 4, and 1 with hamming distance (HD) 0, 1, 2, 3 and 4 respectively. For 3-variables all classes and their sub-classes are given in **Appendix** I.

**Table 4.1: Shows different sub-classes of CLASS 1**

| Boolean Functions | Decimal value | Hamming distance with respect to Affine Boolean function | No. of Boolean Functions |
|---|---|---|---|
| **00000000** (Affine) | 0 | 0 | 1 |
| 00000010 | 2 | 1 | 4 |
| 00100000 | 32 | | |
| 00001000 | 8 | | |
| 00000100 | 4 | | |
| 00100010 | 34 | 2 | 6 |
| 00001010 | 10 | | |
| 00101000 | 40 | | |
| 00001100 | 12 | | |
| 00000110 | 6 | | |
| 00100100 | 36 | | |
| 00101010 | 42 | 3 | 4 |
| 00001110 | 14 | | |
| 00101100 | 44 | | |
| 00100110 | 38 | | |
| 00101110 | 46 | 4 | 1 |

### 4.2 XOR Operation in Classes

Let $a = a_{2^n}, a_{2^n-1}, \ldots \ldots a_1$ and $b = b_{2^n}, b_{2^n-1}, \ldots b_1$ are two Boolean functions of same CLASS of n-variable. The **XOR** operation of all the classes when arranged in a table gives the Boolean functions and all belonging to Class 1 of n-variable as $(a+k) \oplus (b+k) = (a \oplus b) + (k \oplus k) = (a \oplus b)$. Where, the **XOR** operation of a and b is defined as $a \oplus b = (a_{2^n} \oplus b_{2^n} \; a_{2^n-1} \oplus b_{2^n-1} \ldots a_1 \oplus b_1)$.

**Illustration:** Suppose we want the **XOR** operation of $(44)_{10} = (00101100)_2$ and $(34)_{10} = (00100010)_2$ both are belongs to Class 1 of 3-variables. $44 \oplus 34 = (00101100) \oplus (00100010) = (00001110) = 14$. Table 4.2 is constructed for all classes of n-variable Boolean functions that contain only the XOR values of all the functions in a class. The functions are arranged in ascending order in both rows and columns of the table. It can be proved that the content of each table remain invariant under the XOR operation and the decimal values of the table content are same as in **Class1 (ref. Appendix II).**

**Table 4.2: Shows XOR values of Class 1 Boolean functions of 3-variable**

| | CLASS 1 | | | | | | | | | | | | | | |
|---|---|---|---|---|---|---|---|---|---|---|---|---|---|---|---|
| XOR | 0 | 2 | 4 | 6 | 8 | 10 | 12 | 14 | 32 | 34 | 36 | 38 | 40 | 42 | 44 | 46 |
| **0** | 0 | 2 | 4 | 6 | 8 | 10 | 12 | 14 | 32 | 34 | 36 | 38 | 40 | 42 | 44 | 46 |
| **2** | 2 | 0 | 6 | 4 | 10 | 8 | 14 | 12 | 34 | 32 | 38 | 36 | 42 | 40 | 46 | 44 |
| **4** | 4 | 6 | 0 | 2 | 12 | 14 | 8 | 10 | 36 | 38 | 32 | 34 | 44 | 46 | 40 | 42 |
| **6** | 6 | 4 | 2 | 0 | 14 | 12 | 10 | 8 | 38 | 36 | 34 | 32 | 46 | 44 | 42 | 40 |
| **8** | 8 | 10 | 12 | 14 | 0 | 2 | 4 | 6 | 40 | 42 | 44 | 46 | 32 | 34 | 36 | 38 |
| **10** | 10 | 8 | 14 | 12 | 2 | 0 | 6 | 4 | 42 | 40 | 46 | 44 | 34 | 32 | 38 | 36 |
| **12** | 12 | 14 | 8 | 10 | 4 | 6 | 0 | 2 | 44 | 46 | 40 | 42 | 36 | 38 | 32 | 34 |
| **14** | 14 | 12 | 10 | 8 | 6 | 4 | 2 | 0 | 46 | 44 | 42 | 40 | 38 | 36 | 34 | 32 |
| **32** | 32 | 34 | 36 | 38 | 40 | 42 | 44 | 46 | 0 | 2 | 4 | 6 | 8 | 10 | 12 | 14 |
| **34** | 34 | 32 | 38 | 36 | 42 | 40 | 46 | 44 | 2 | 0 | 6 | 4 | 10 | 8 | 14 | 12 |
| **36** | 36 | 38 | 32 | 34 | 44 | 46 | 40 | 42 | 4 | 6 | 0 | 2 | 12 | 14 | 8 | 10 |
| **38** | 38 | 36 | 34 | 32 | 46 | 44 | 42 | 40 | 6 | 4 | 2 | 0 | 14 | 12 | 10 | 8 |
| **40** | 40 | 42 | 44 | 46 | 32 | 34 | 36 | 38 | 8 | 10 | 12 | 14 | 0 | 2 | 4 | 6 |
| **42** | 42 | 40 | 46 | 44 | 34 | 32 | 38 | 36 | 10 | 8 | 14 | 12 | 2 | 0 | 6 | 4 |
| **44** | 44 | 46 | 40 | 42 | 36 | 38 | 32 | 34 | 12 | 14 | 8 | 10 | 4 | 6 | 0 | 2 |
| **46** | 46 | 44 | 42 | 40 | 38 | 36 | 34 | 32 | 14 | 12 | 10 | 8 | 6 | 4 | 2 | 0 |

### 4.3 CVT Operation in Classes

Let $a = a_k, a_{k-1}, ..., a_1$ and $b = b_k, b_{k-1}, ..., b_1$ are two Boolean functions in Class. Then the Carry Value Transform (**CVT**) operation of a and b is defined in[10] as $CVT(a,b) = a_k \wedge b_k, a_{k-1} \wedge b_{k-1}, ..., a_1 \wedge b_1, 0$.
Carry Value Transformation (CVT) is a kind of representation of n-variable Boolean functions and is used to produce many interesting patterns [10]. We observe that some interesting **patterns** are generated which **are invariant** for all classes of n-variable Boolean functions under CVT operation.

**Illustration:** The **CVT** operation of $(44)_{10} = (00101100)_2$ and $(34)_{10} = (00100010)_2$ both are belongs to Class 1 of 3-variable and the $CVT(44,34) = CVT(00101100, 00100010) = (00100000) = 64$. The patterns for class 1 functions using CVT operation is shown in table 4.3 and others are in **Appendix II**.

**Table 4.3: Shows CVT Pattern of Class 1 Boolean functions in 3-variables**

| | CLASS 1 | | | | | | | | | | | | | | |
|---|---|---|---|---|---|---|---|---|---|---|---|---|---|---|---|
| CVT | 0 | 2 | 4 | 6 | 8 | 10 | 12 | 14 | 32 | 34 | 36 | 38 | 40 | 42 | 44 | 46 |
| **0** | 0 | 0 | 0 | 0 | 0 | 0 | 0 | 0 | 0 | 0 | 0 | 0 | 0 | 0 | 0 | 0 |
| **2** | 0 | 4 | 0 | 4 | 0 | 4 | 0 | 4 | 0 | 4 | 0 | 4 | 0 | 4 | 0 | 4 |
| **4** | 0 | 0 | 8 | 8 | 0 | 0 | 8 | 8 | 0 | 0 | 8 | 8 | 0 | 0 | 8 | 8 |
| **6** | 0 | 4 | 8 | 12 | 0 | 4 | 8 | 12 | 0 | 4 | 8 | 12 | 0 | 4 | 8 | 12 |
| **8** | 0 | 0 | 0 | 0 | 16 | 16 | 16 | 16 | 0 | 0 | 0 | 0 | 16 | 16 | 16 | 16 |
| **10** | 0 | 4 | 0 | 4 | 16 | 20 | 16 | 20 | 0 | 4 | 0 | 4 | 16 | 20 | 16 | 20 |
| **12** | 0 | 0 | 8 | 8 | 16 | 16 | 24 | 24 | 0 | 0 | 8 | 8 | 16 | 16 | 24 | 24 |
| **14** | 0 | 4 | 8 | 12 | 16 | 20 | 24 | 28 | 0 | 4 | 8 | 12 | 16 | 20 | 24 | 28 |
| **32** | 0 | 0 | 0 | 0 | 0 | 0 | 0 | 0 | 64 | 64 | 64 | 64 | 64 | 64 | 64 | 64 |
| **34** | 0 | 4 | 0 | 4 | 0 | 4 | 0 | 4 | 64 | 68 | 64 | 68 | 64 | 68 | 64 | 68 |
| **36** | 0 | 0 | 8 | 8 | 0 | 0 | 8 | 8 | 64 | 64 | 72 | 72 | 64 | 64 | 72 | 72 |
| **38** | 0 | 4 | 8 | 12 | 0 | 4 | 8 | 12 | 64 | 68 | 72 | 76 | 64 | 68 | 72 | 76 |
| **40** | 0 | 0 | 0 | 0 | 16 | 16 | 16 | 16 | 64 | 64 | 64 | 64 | 80 | 80 | 80 | 80 |
| **42** | 0 | 4 | 0 | 4 | 16 | 20 | 16 | 20 | 64 | 68 | 64 | 68 | 80 | 84 | 80 | 84 |
| **44** | 0 | 0 | 8 | 8 | 16 | 16 | 24 | 24 | 64 | 64 | 72 | 72 | 80 | 80 | 88 | 88 |
| **46** | 0 | 4 | 8 | 12 | 16 | 20 | 24 | 28 | 64 | 68 | 72 | 76 | 80 | 84 | 88 | 92 |

## 5. Conclusion and Future Endeavours

In this paper, the set of all possible n-variable Boolean functions are divided into disjoint equivalence classes of same size. The beauty of this classification is that each class contains exactly one affine Boolean function and all other non-affine functions. This classification where affine functions are uniformly distributed is achieved through two different procedures. The characteristics of the classes and the behaviour of those classes with respect to Hamming distance, XOR and CVT operations are studied.

**Acknowledgement:** The authors are grateful to *Prof. Birendra Kumar Nayak of Utkal University and Mr. Sk. Sarif Hassan of Institute of Mathematics and Applications, Bhubaneswar* for their valuable suggestions.

## References

[1] S. W Golomb (1959), " On the classification of Boolean functions", *IRE transactions on circuit theory*, *Vol. 06, Issue. 05, pp.176- 186 .*

[2] D. Slepian (1954), "On the number of symmetry types of Boolean functions of n variables", Society for industrial and *Mathematics", Vol. 5, No. 2, pp. 185-193.*

[3] M. A. Harrison (1964), "On the classification of Boolean functions by the general linear and affine groups*",Journal of the Society for Industrial and Applied Mathematics, Vol. 12, No. 2, pp. 285-299*

[4] V. P. Correia, A. I. Reis (2001), "Classifying n-Input Boolean Functions*.", IBERCHIP, pp.58-66*

[5] An. Braeken, Y. Borissov, S. Nikova, B. Preneel (2005), "Classification of Boolean Functions of 6 Variables or Less with Respect to Cryptographic Properties", *ICALP'05 Proceedings of the 32nd international conference on Automata, Languages and Programming,Springer-Verlag Berlin, Heidelberg , pp. 324-334*

[6] P. Pal Choudhury, S. Sahoo, M. Chakarborty, S. K Bhandari, A. Pal(2009), *"*Investigation of the Global Dynamics of Cellular Automata Using Boolean Derivatives ", Computers and Mathematics with Applications*, Elsevier, 57, pp. 1337-1351 .*

[7] P. Pal Choudhury, S. Sahoo, M. Chakraborty (2011) ,"Characterization of the evolution of Non-linear Uniform Cellular Automata in the light of Deviant States", *Mathematics and Mathematical Sciences*, Vol. 2011, Article ID 605098.

[8] S. Sahoo, P. Pal Choudhury, M. Chakraborty (2010), "Characterization of any Non-linear Boolean function Using A Set of Linear Operators", *Journal of Orissa Mathematical Society, Vol. 2, No. 1&2, pp. 111-133*

[9] B. K Nayak, S. Sahoo, S. Biswal,"Cellular Automata Rules and Linear Numbers", *arxiv.org*/pdf /1204.3999.

[10] P. Pal Choudhury, S. Sahoo, B. K Nayak, Sk. Sarif Hassan(2010), "Theory of Carry Value Transformation (CVT) and its Application in Fractal formation", *Global Journal of Computer Science and Technology, Vol. 10, Issue 14, Version 1.0, pp. 98-107 .*

# Appendix I

*Following table shows the Classes and Sub-classes of 3-variable Boolean Functions.* **Abbreviations:** *BF-Boolean Function, DV-Decimal Value, HD-Hamming Distance, No.BF-Number of Boolean Functions.*

| \multicolumn{4}{c}{**Class 2**} | \multicolumn{4}{c}{**Class 3**} | \multicolumn{4}{c}{**Class 4**} |
|---|---|---|---|---|---|---|---|---|---|---|---|
| BF | DV | HD | No.BF | BF | DV | HD | No.BF | BF | DV | HD | No.BF |
| **10101010** | **170** | 0 | 1 | **01100110** | **204** | 0 | 1 | **01100110** | **102** | 0 | 1 |
| 10100010 | 162 |   |   | 01100010 | 200 |   |   | 01100010 | 98 |   |   |
| 10101000 | 168 | 1 | 4 | 01101110 | 206 | 1 | 4 | 01101110 | 110 | 1 | 4 |
| 10001010 | 138 |   |   | 01100100 | 236 |   |   | 01100100 | 100 |   |   |
| 10101110 | 174 |   |   | 01000110 | 196 |   |   | 01000110 | 70 |   |   |
| 10100000 | 160 |   |   | 11000000 | 192 |   |   | 01100000 | 96 |   |   |
| 10000010 | 130 |   |   | 11001010 | 202 |   |   | 01000010 | 66 |   |   |
| 10001000 | 136 | 2 | 6 | 11101000 | 232 | 2 | 6 | 01101010 | 106 | 2 | 6 |
| 10101100 | 172 |   |   | 11101110 | 238 |   |   | 01101100 | 108 |   |   |
| 10001110 | 142 |   |   | 11000110 | 198 |   |   | 01001110 | 78 |   |   |
| 10100110 | 166 |   |   | 11100100 | 228 |   |   | 01000100 | 68 |   |   |
| 10000000 | 128 |   |   | 11000010 | 194 |   |   | 01000000 | 64 |   |   |
| 10001100 | 140 | 3 | 4 | 11100000 | 224 | 3 | 4 | 01101000 | 104 | 3 | 4 |
| 10100100 | 164 |   |   | 11101010 | 234 |   |   | 01001010 | 74 |   |   |
| 10000110 | 134 |   |   | 11100110 | 230 |   |   | 01001100 | 76 |   |   |
| 10000100 | 132 | 4 | 1 | 11100010 | 226 | 4 | 1 | 01001000 | 72 | 4 | 1 |
| \multicolumn{4}{c}{**Class 5**} | \multicolumn{4}{c}{**Class 6**} | \multicolumn{4}{c}{**Class 7**} |
| BF | DV | HD | No.BF | BF | DV | HD | No.BF | BF | DV | HD | No.BF |
| **1111000** | **240** | 0 | 1 | **01011010** | **90** | 0 | 1 | **0011110** | **60** | 0 | 1 |
| 11110010 | 242 |   |   | 01010010 | 82 |   |   | 00111000 | 56 |   |   |
| 11010000 | 208 | 1 | 4 | 01011000 | 88 | 1 | 4 | 00111110 | 62 | 1 | 4 |
| 11111000 | 248 |   |   | 01111010 | 122 |   |   | 00011100 | 28 |   |   |
| 11110100 | 244 |   |   | 01011110 | 94 |   |   | 00110100 | 52 |   |   |
| 11010010 | 210 |   |   | 01010000 | 80 |   |   | 00110000 | 48 |   |   |
| 11111010 | 250 |   |   | 01110010 | 114 |   |   | 00111010 | 58 |   |   |
| 11011000 | 216 | 2 | 6 | 01111000 | 120 | 2 | 6 | 00011000 | 24 | 2 | 6 |
| 11111100 | 252 |   |   | 01011100 | 92 |   |   | 00011110 | 30 |   |   |
| 11110110 | 246 |   |   | 01111110 | 126 |   |   | 00110110 | 54 |   |   |
| 11010100 | 212 |   |   | 01010110 | 86 |   |   | 00010100 | 20 |   |   |
| 11011010 | 218 |   |   | 01110000 | 112 |   |   | 00110010 | 50 |   |   |
| 11111110 | 254 | 3 | 4 | 01111100 | 124 | 3 | 4 | 00010000 | 16 | 3 | 4 |
| 11011100 | 220 |   |   | 01010100 | 84 |   |   | 00011010 | 26 |   |   |
| 11010110 | 214 |   |   | 01110110 | 118 |   |   | 00010110 | 22 |   |   |
| 11011110 | 222 | 4 | 1 | 01110100 | 116 | 4 | 1 | 00010010 | 18 | 4 | 1 |

| Class 8 | | | | Class 9 | | | | Class 10 | | | |
|---|---|---|---|---|---|---|---|---|---|---|---|
| BF | DV | HD | No.BF | BF | DV | HD | No.BF | BF | DV | HD | No.BF |
| **10010110** | **150** | 0 | 1 | **11111111** | **255** | 0 | 1 | **01010101** | **85** | 0 | 1 |
| 10010010 | 146 |   |   | 11111101 | 253 |   |   | 01011101 | 93 |   |   |
| 10011110 | 158 | 1 | 4 | 11011111 | 223 | 1 | 4 | 01010111 | 87 | 1 | 4 |
| 10010100 | 148 |   |   | 11110111 | 247 |   |   | 01110101 | 117 |   |   |
| 10110110 | 182 |   |   | 11111011 | 251 |   |   | 01010001 | 81 |   |   |
| 10010000 | 144 |   |   | 11011101 | 221 |   |   | 01011111 | 95 |   |   |
| 10110010 | 178 |   |   | 11110101 | 245 |   |   | 01111101 | 125 |   |   |
| 10011010 | 154 | 2 | 6 | 11010111 | 215 | 2 | 6 | 01110111 | 119 | 2 | 6 |
| 10011100 | 156 |   |   | 11110011 | 243 |   |   | 01010011 | 83 |   |   |
| 10111110 | 190 |   |   | 11111001 | 249 |   |   | 01110001 | 113 |   |   |
| 10110100 | 180 |   |   | 11011011 | 219 |   |   | 01011001 | 89 |   |   |
| 10110000 | 176 |   |   | 11010101 | 213 |   |   | 01111111 | 127 |   |   |
| 10011000 | 152 | 3 | 4 | 11110001 | 241 | 3 | 4 | 01110011 | 115 | 3 | 4 |
| 10111010 | 186 |   |   | 11010011 | 211 |   |   | 01011011 | 91 |   |   |
| 10111100 | 188 |   |   | 11011001 | 217 |   |   | 01111001 | 121 |   |   |
| 10111000 | 184 | 4 | 1 | 11010001 | 209 | 4 | 1 | 01111011 | 123 | 4 | 1 |
| Class 11 | | | | Class 12 | | | | Class 13 | | | |
| BF | DV | HD | No.BF | BF | DV | HD | No.BF | BF | DV | HD | No.BF |
| **00110011** | **51** | 0 | 1 | **10011001** | **153** | 0 | 1 | **00001111** | **15** | 0 | 1 |
| 00110111 | 55 |   |   | 10011101 | 157 |   |   | 00001101 | 13 |   |   |
| 00110001 | 49 | 1 | 4 | 10010001 | 145 | 1 | 4 | 00101111 | 47 | 1 | 4 |
| 00010011 | 19 |   |   | 10011011 | 155 |   |   | 00000111 | 7 |   |   |
| 00111011 | 59 |   |   | 10111001 | 185 |   |   | 00001011 | 11 |   |   |
| 00111111 | 63 |   |   | 10011111 | 159 |   |   | 00101101 | 45 |   |   |
| 00110101 | 53 |   |   | 10111101 | 189 |   |   | 00000101 | 5 |   |   |
| 00010111 | 23 | 2 | 6 | 10010101 | 149 | 2 | 6 | 00100111 | 39 | 2 | 6 |
| 00010001 | 17 |   |   | 10010011 | 147 |   |   | 00000011 | 3 |   |   |
| 00111001 | 57 |   |   | 10110001 | 177 |   |   | 00001001 | 9 |   |   |
| 00011011 | 27 |   |   | 10111011 | 187 |   |   | 00101011 | 43 |   |   |
| 00111101 | 61 |   |   | 10111111 | 191 |   |   | 00100101 | 37 |   |   |
| 00011111 | 31 | 3 | 4 | 10010111 | 151 | 3 | 4 | 00000001 | 1 | 3 | 4 |
| 00010101 | 21 |   |   | 10110101 | 181 |   |   | 00100011 | 35 |   |   |
| 00011001 | 25 |   |   | 10110011 | 179 |   |   | 00101001 | 41 |   |   |
| 00011101 | 29 | 4 | 1 | 10110111 | 183 | 4 | 1 | 00100001 | 33 | 4 | 1 |

| Class 14 | | | | Class 15 | | | | Class 16 | | | |
|---|---|---|---|---|---|---|---|---|---|---|---|
| BF | DV | HD | No.BF | BF | DV | HD | No.BF | BF | DV | HD | No.BF |
| **10100101** | **165** | 0 | 1 | **11000011** | **195** | 0 | 1 | **01101001** | **105** | 0 | 1 |
| 10101101 | 173 |   |   | 11000111 | 199 |   |   | 01101101 | 109 |   |   |
| 10100111 | 167 | 1 | 4 | 11000001 | 193 | 1 | 4 | 01100001 | 97 | 1 | 4 |
| 10000101 | 133 |   |   | 11100011 | 227 |   |   | 01101011 | 107 |   |   |
| 10100001 | 161 |   |   | 11001011 | 203 |   |   | 01001001 | 73 |   |   |
| 10101111 | 175 |   |   | 11001111 | 207 |   |   | 01101111 | 111 |   |   |
| 10001101 | 141 |   |   | 11000101 | 197 |   |   | 01001101 | 77 |   |   |
| 10000111 | 135 | 2 | 6 | 11100111 | 231 | 2 | 6 | 01100101 | 101 | 2 | 6 |
| 10100011 | 163 |   |   | 11100001 | 225 |   |   | 01100011 | 99 |   |   |
| 10000001 | 129 |   |   | 11001001 | 201 |   |   | 01000001 | 65 |   |   |
| 10101001 | 169 |   |   | 11101011 | 235 |   |   | 01001011 | 75 |   |   |
| 10001111 | 143 |   |   | 11001101 | 205 |   |   | 01001111 | 79 |   |   |
| 10000011 | 131 | 3 | 4 | 11101111 | 239 | 3 | 4 | 01100111 | 103 | 3 | 4 |
| 10101011 | 171 |   |   | 11100101 | 229 |   |   | 01000101 | 69 |   |   |
| 10001001 | 137 |   |   | 11101001 | 233 |   |   | 01000011 | 67 |   |   |
| 10001011 | 139 | 4 | 1 | 11101101 | 237 | 4 | 1 | 01000111 | 71 | 4 | 1 |

# Appendix II

*Following table shows the XOR operation values of some classes of 3-variable Boolean Functions.*

| CLASS 1 | | | | | | | | | | | | | | | | |
|---|---|---|---|---|---|---|---|---|---|---|---|---|---|---|---|---|
| *XOR* | **0** | **2** | **4** | **6** | **8** | **10** | **12** | **14** | **32** | **34** | **36** | **38** | **40** | **42** | **44** | **46** |
| **0** | 0 | 2 | 4 | 6 | 8 | 10 | 12 | 14 | 32 | 34 | 36 | 38 | 40 | 42 | 44 | 46 |
| **2** | 2 | 0 | 6 | 4 | 10 | 8 | 14 | 12 | 34 | 32 | 38 | 36 | 42 | 40 | 46 | 44 |
| **4** | 4 | 6 | 0 | 2 | 12 | 14 | 8 | 10 | 36 | 38 | 32 | 34 | 44 | 46 | 40 | 42 |
| **6** | 6 | 4 | 2 | 0 | 14 | 12 | 10 | 8 | 38 | 36 | 34 | 32 | 46 | 44 | 42 | 40 |
| **8** | 8 | 10 | 12 | 14 | 0 | 2 | 4 | 6 | 40 | 42 | 44 | 46 | 32 | 34 | 36 | 38 |
| **10** | 10 | 8 | 14 | 12 | 2 | 0 | 6 | 4 | 42 | 40 | 46 | 44 | 34 | 32 | 38 | 36 |
| **12** | 12 | 14 | 8 | 10 | 4 | 6 | 0 | 2 | 44 | 46 | 40 | 42 | 36 | 38 | 32 | 34 |
| **14** | 14 | 12 | 10 | 8 | 6 | 4 | 2 | 0 | 46 | 44 | 42 | 40 | 38 | 36 | 34 | 32 |
| **32** | 32 | 34 | 36 | 38 | 40 | 42 | 44 | 46 | 0 | 2 | 4 | 6 | 8 | 10 | 12 | 14 |
| **34** | 34 | 32 | 38 | 36 | 42 | 40 | 46 | 44 | 2 | 0 | 6 | 4 | 10 | 8 | 14 | 12 |
| **36** | 36 | 38 | 32 | 34 | 44 | 46 | 40 | 42 | 4 | 6 | 0 | 2 | 12 | 14 | 8 | 10 |
| **38** | 38 | 36 | 34 | 32 | 46 | 44 | 42 | 40 | 6 | 4 | 2 | 0 | 14 | 12 | 10 | 8 |
| **40** | 40 | 42 | 44 | 46 | 32 | 34 | 36 | 38 | 8 | 10 | 12 | 14 | 0 | 2 | 4 | 6 |
| **42** | 42 | 40 | 46 | 44 | 34 | 32 | 38 | 36 | 10 | 8 | 14 | 12 | 2 | 0 | 6 | 4 |
| **44** | 44 | 46 | 40 | 42 | 36 | 38 | 32 | 34 | 12 | 14 | 8 | 10 | 4 | 6 | 0 | 2 |
| **46** | 46 | 44 | 42 | 40 | 38 | 36 | 34 | 32 | 14 | 12 | 10 | 8 | 6 | 4 | 2 | 0 |
| **CLASS 2** | | | | | | | | | | | | | | | | |
| *XOR* | **128** | **130** | **132** | **134** | **136** | **138** | **140** | **142** | **160** | **162** | **164** | **166** | **168** | **170** | **172** | **174** |
| **128** | 0 | 2 | 4 | 6 | 8 | 10 | 12 | 14 | 32 | 34 | 36 | 38 | 40 | 42 | 44 | 46 |
| **130** | 2 | 0 | 6 | 4 | 10 | 8 | 14 | 12 | 34 | 32 | 38 | 36 | 42 | 40 | 46 | 44 |
| **132** | 4 | 6 | 0 | 2 | 12 | 14 | 8 | 10 | 36 | 38 | 32 | 34 | 44 | 46 | 40 | 42 |
| **134** | 6 | 4 | 2 | 0 | 14 | 12 | 10 | 8 | 38 | 36 | 34 | 32 | 46 | 44 | 42 | 40 |
| **136** | 8 | 10 | 12 | 14 | 0 | 2 | 4 | 6 | 40 | 42 | 44 | 46 | 32 | 34 | 36 | 38 |
| **138** | 10 | 8 | 14 | 12 | 2 | 0 | 6 | 4 | 42 | 40 | 46 | 44 | 34 | 32 | 38 | 36 |
| **140** | 12 | 14 | 8 | 10 | 4 | 6 | 0 | 2 | 44 | 46 | 40 | 42 | 36 | 38 | 32 | 34 |
| **142** | 14 | 12 | 10 | 8 | 6 | 4 | 2 | 0 | 46 | 44 | 42 | 40 | 38 | 36 | 34 | 32 |
| **160** | 32 | 34 | 36 | 38 | 40 | 42 | 44 | 46 | 0 | 2 | 4 | 6 | 8 | 10 | 12 | 14 |
| **162** | 34 | 32 | 38 | 36 | 42 | 40 | 46 | 44 | 2 | 0 | 6 | 4 | 10 | 8 | 14 | 12 |
| **164** | 36 | 38 | 32 | 34 | 44 | 46 | 40 | 42 | 4 | 6 | 0 | 2 | 12 | 14 | 8 | 10 |
| **166** | 38 | 36 | 34 | 32 | 46 | 44 | 42 | 40 | 6 | 4 | 2 | 0 | 14 | 12 | 10 | 8 |
| **168** | 40 | 42 | 44 | 46 | 32 | 34 | 36 | 38 | 8 | 10 | 12 | 14 | 0 | 2 | 4 | 6 |
| **170** | 42 | 40 | 46 | 44 | 34 | 32 | 38 | 36 | 10 | 8 | 14 | 12 | 2 | 0 | 6 | 4 |
| **172** | 44 | 46 | 40 | 42 | 36 | 38 | 32 | 34 | 12 | 14 | 8 | 10 | 4 | 6 | 0 | 2 |
| **174** | 46 | 44 | 42 | 40 | 38 | 36 | 34 | 32 | 14 | 12 | 10 | 8 | 6 | 4 | 2 | 0 |
| **CLASS 3** | | | | | | | | | | | | | | | | |
| *XOR* | **192** | **194** | **196** | **198** | **200** | **202** | **204** | **206** | **224** | **226** | **228** | **230** | **232** | **234** | **236** | **238** |
| **192** | 0 | 2 | 4 | 6 | 8 | 10 | 12 | 14 | 32 | 34 | 36 | 38 | 40 | 42 | 44 | 46 |
| **194** | 2 | 0 | 6 | 4 | 10 | 8 | 14 | 12 | 34 | 32 | 38 | 36 | 42 | 40 | 46 | 44 |
| **196** | 4 | 6 | 0 | 2 | 12 | 14 | 8 | 10 | 36 | 38 | 32 | 34 | 44 | 46 | 40 | 42 |
| **198** | 6 | 4 | 2 | 0 | 14 | 12 | 10 | 8 | 38 | 36 | 34 | 32 | 46 | 44 | 42 | 40 |
| **200** | 8 | 10 | 12 | 14 | 0 | 2 | 4 | 6 | 40 | 42 | 44 | 46 | 32 | 34 | 36 | 38 |
| **202** | 10 | 8 | 14 | 12 | 2 | 0 | 6 | 4 | 42 | 40 | 46 | 44 | 34 | 32 | 38 | 36 |
| **204** | 12 | 14 | 8 | 10 | 4 | 6 | 0 | 2 | 44 | 46 | 40 | 42 | 36 | 38 | 32 | 34 |
| **206** | 14 | 12 | 10 | 8 | 6 | 4 | 2 | 0 | 46 | 44 | 42 | 40 | 38 | 36 | 34 | 32 |
| **224** | 32 | 34 | 36 | 38 | 40 | 42 | 44 | 46 | 0 | 2 | 4 | 6 | 8 | 10 | 12 | 14 |
| **226** | 34 | 32 | 38 | 36 | 42 | 40 | 46 | 44 | 2 | 0 | 6 | 4 | 10 | 8 | 14 | 12 |
| **228** | 36 | 38 | 32 | 34 | 44 | 46 | 40 | 42 | 4 | 6 | 0 | 2 | 12 | 14 | 8 | 10 |
| **230** | 38 | 36 | 34 | 32 | 46 | 44 | 42 | 40 | 6 | 4 | 2 | 0 | 14 | 12 | 10 | 8 |
| **232** | 40 | 42 | 44 | 46 | 32 | 34 | 36 | 38 | 8 | 10 | 12 | 14 | 0 | 2 | 4 | 6 |
| **234** | 42 | 40 | 46 | 44 | 34 | 32 | 38 | 36 | 10 | 8 | 14 | 12 | 2 | 0 | 6 | 4 |
| **236** | 44 | 46 | 40 | 42 | 36 | 38 | 32 | 34 | 12 | 14 | 8 | 10 | 4 | 6 | 0 | 2 |
| **238** | 46 | 44 | 42 | 40 | 38 | 36 | 34 | 32 | 14 | 12 | 10 | 8 | 6 | 4 | 2 | 0 |

*Following table shows the CVT operation values of some classes of 3-variable Boolean Functions.*

| CLASS 1 | | | | | | | | | | | | | | | | |
|---|---|---|---|---|---|---|---|---|---|---|---|---|---|---|---|---|
| CVT | 0 | 2 | 4 | 6 | 8 | 10 | 12 | 14 | 32 | 34 | 36 | 38 | 40 | 42 | 44 | 46 |
| 0 | 0 | 0 | 0 | 0 | 0 | 0 | 0 | 0 | 0 | 0 | 0 | 0 | 0 | 0 | 0 | 0 |
| 2 | 0 | 4 | 0 | 4 | 0 | 4 | 0 | 4 | 0 | 4 | 0 | 4 | 0 | 4 | 0 | 4 |
| 4 | 0 | 0 | 8 | 8 | 0 | 0 | 8 | 8 | 0 | 0 | 8 | 8 | 0 | 0 | 8 | 8 |
| 6 | 0 | 4 | 8 | 12 | 0 | 4 | 8 | 12 | 0 | 4 | 8 | 12 | 0 | 4 | 8 | 12 |
| 8 | 0 | 0 | 0 | 0 | 16 | 16 | 16 | 16 | 0 | 0 | 0 | 0 | 16 | 16 | 16 | 16 |
| 10 | 0 | 4 | 0 | 4 | 16 | 20 | 16 | 20 | 0 | 4 | 0 | 4 | 16 | 20 | 16 | 20 |
| 12 | 0 | 0 | 8 | 8 | 16 | 16 | 24 | 24 | 0 | 0 | 8 | 8 | 16 | 16 | 24 | 24 |
| 14 | 0 | 4 | 8 | 12 | 16 | 20 | 24 | 28 | 0 | 4 | 8 | 12 | 16 | 20 | 24 | 28 |
| 32 | 0 | 0 | 0 | 0 | 0 | 0 | 0 | 0 | 64 | 64 | 64 | 64 | 64 | 64 | 64 | 64 |
| 34 | 0 | 4 | 0 | 4 | 0 | 4 | 0 | 4 | 64 | 68 | 64 | 68 | 64 | 68 | 64 | 68 |
| 36 | 0 | 0 | 8 | 8 | 0 | 0 | 8 | 8 | 64 | 64 | 72 | 72 | 64 | 64 | 72 | 72 |
| 38 | 0 | 4 | 8 | 12 | 0 | 4 | 8 | 12 | 64 | 68 | 72 | 76 | 64 | 68 | 72 | 76 |
| 40 | 0 | 0 | 0 | 0 | 16 | 16 | 16 | 16 | 64 | 64 | 64 | 64 | 80 | 80 | 80 | 80 |
| 42 | 0 | 4 | 0 | 4 | 16 | 20 | 16 | 20 | 64 | 68 | 64 | 68 | 80 | 84 | 80 | 84 |
| 44 | 0 | 0 | 8 | 8 | 16 | 16 | 24 | 24 | 64 | 64 | 72 | 72 | 80 | 80 | 88 | 88 |
| 46 | 0 | 4 | 8 | 12 | 16 | 20 | 24 | 28 | 64 | 68 | 72 | 76 | 80 | 84 | 88 | 92 |

| CLASS 2 | | | | | | | | | | | | | | | | |
|---|---|---|---|---|---|---|---|---|---|---|---|---|---|---|---|---|
| CVT | 128 | 130 | 132 | 134 | 136 | 138 | 140 | 142 | 160 | 162 | 164 | 166 | 168 | 170 | 172 | 174 |
| 128 | 256 | 256 | 256 | 256 | 256 | 256 | 256 | 256 | 256 | 256 | 256 | 256 | 256 | 256 | 256 | 256 |
| 130 | 256 | 260 | 256 | 260 | 256 | 260 | 256 | 260 | 256 | 260 | 256 | 260 | 256 | 260 | 256 | 260 |
| 132 | 256 | 256 | 264 | 264 | 256 | 256 | 264 | 264 | 256 | 256 | 264 | 264 | 256 | 256 | 264 | 264 |
| 134 | 256 | 260 | 264 | 268 | 256 | 260 | 264 | 268 | 256 | 260 | 264 | 268 | 256 | 260 | 264 | 268 |
| 136 | 256 | 256 | 256 | 256 | 272 | 272 | 272 | 272 | 256 | 256 | 256 | 256 | 272 | 272 | 272 | 272 |
| 138 | 256 | 260 | 256 | 260 | 272 | 276 | 272 | 276 | 256 | 260 | 256 | 260 | 272 | 276 | 272 | 276 |
| 140 | 256 | 256 | 264 | 264 | 272 | 272 | 280 | 280 | 256 | 256 | 264 | 264 | 272 | 272 | 280 | 280 |
| 142 | 256 | 260 | 264 | 268 | 272 | 276 | 280 | 284 | 256 | 260 | 264 | 268 | 272 | 276 | 280 | 284 |
| 160 | 256 | 256 | 256 | 256 | 256 | 256 | 256 | 256 | 320 | 320 | 320 | 320 | 320 | 320 | 320 | 320 |
| 162 | 256 | 260 | 256 | 260 | 256 | 260 | 256 | 260 | 320 | 324 | 320 | 324 | 320 | 324 | 320 | 324 |
| 164 | 256 | 256 | 264 | 264 | 256 | 256 | 264 | 264 | 320 | 320 | 328 | 328 | 320 | 320 | 328 | 328 |
| 166 | 256 | 260 | 264 | 268 | 256 | 260 | 264 | 268 | 320 | 324 | 328 | 332 | 320 | 324 | 328 | 332 |
| 168 | 256 | 256 | 256 | 256 | 272 | 272 | 272 | 272 | 320 | 320 | 320 | 320 | 336 | 336 | 336 | 336 |
| 170 | 256 | 260 | 256 | 260 | 272 | 276 | 272 | 276 | 320 | 324 | 320 | 324 | 336 | 340 | 336 | 340 |
| 172 | 256 | 256 | 264 | 264 | 272 | 272 | 280 | 280 | 320 | 320 | 328 | 328 | 336 | 336 | 344 | 344 |
| 174 | 256 | 260 | 264 | 268 | 272 | 276 | 280 | 284 | 320 | 324 | 328 | 332 | 336 | 340 | 344 | 348 |

| CLASS 3 | | | | | | | | | | | | | | | | |
|---|---|---|---|---|---|---|---|---|---|---|---|---|---|---|---|---|
| CVT | 192 | 194 | 196 | 198 | 200 | 202 | 204 | 206 | 224 | 226 | 228 | 230 | 232 | 234 | 236 | 238 |
| 192 | | | | | | | | | | | | | | | | |
| 194 | | | | | | | | | | | | | | | | |
| 196 | | | | | | | | | | | | | | | | |
| 198 | | | | | | | | | | | | | | | | |
| 200 | | | | | | | | | | | | | | | | |
| 202 | | | | | | | | | | | | | | | | |
| 204 | | | | | | | | | | | | | | | | |
| 206 | | | | | | | | | | | | | | | | |
| 224 | | | | | | | | | | | | | | | | |
| 226 | | | | | | | | | | | | | | | | |
| 228 | | | | | | | | | | | | | | | | |
| 230 | | | | | | | | | | | | | | | | |
| 232 | | | | | | | | | | | | | | | | |
| 234 | | | | | | | | | | | | | | | | |
| 236 | | | | | | | | | | | | | | | | |
| 238 | | | | | | | | | | | | | | | | |